\documentclass[preprint,showpacs,12pt,nofootinbib]{revtex4}

\usepackage{amsmath,amssymb,amsfonts}
\usepackage{graphicx}% Include figure files
\usepackage{hhline}
\usepackage{bm}
\usepackage{amsmath}
\usepackage{epsfig}
\usepackage{graphicx}

\usepackage{tabularx}
\usepackage{lscape}
\usepackage{rotating}
\usepackage{pstricks}

\usepackage{ulem}

\begin{document}
\title{Neural Network predictions of inclusive electron-nucleus cross sections}

\author{O. Al Hammal}
\affiliation{IPSA-DRII,  63 boulevard de Brandebourg, 94200 Ivry-sur-Seine, France}
\author{M. Martini}
\affiliation{IPSA-DRII,  63 boulevard de Brandebourg, 94200 Ivry-sur-Seine, France}
\affiliation{Sorbonne Universit\'e, Universit\'e Paris Diderot, CNRS/IN2P3, Laboratoire
de Physique Nucl\'eaire et de Hautes Energies (LPNHE), Paris, France}
\author{J. Frontera-Pons}
\affiliation{IPSA-DRII,  63 boulevard de Brandebourg, 94200 Ivry-sur-Seine, France}
\affiliation{Laboratoire AIM, CEA, CNRS, Universit\'e Paris-Saclay, Universit\'e de Paris, 91191 Gif-sur-Yvette, France}
\author{T. H.  Nguyen}
\affiliation{IPSA,  63 boulevard de Brandebourg, 94200 Ivry-sur-Seine, France}
\author{R. P\'erez-Ramos}
\affiliation{IPSA-DRII,  63 boulevard de Brandebourg, 94200 Ivry-sur-Seine, France}
\affiliation{Sorbonne Universit\'e, Universit\'e Paris Diderot, CNRS/IN2P3, Laboratoire
de Physique Th\'eorique et de Hautes Energies (LPTHE), Paris, France}

\begin{abstract}
We investigate whether a neural network approach can reproduce and predict the electron-nucleus cross sections in the kinematical domain of present and future accelerator-based neutrino oscillation experiments.   
For this purpose, we consider the large amount of data 
available to the community via the web-page ``Quasielastic Electron Nucleus scattering archive'', and use a residual, fully connected feedforward neural network.
We illustrate the training performances of the neural network by comparing its results with experimental data for the electron double-differential cross section on carbon. The agreement between predictions and data is remarkable from quasielastic to deep-inelastic scattering. To test the predicting power of the neural network we consider the numerous kinematical conditions for which experimental cross sections on calcium are available. Furthermore, we show the predictions of the electron scattering cross sections on oxygen, argon, and titanium: nuclei of particular interest in the context of present and future accelerator-based neutrino oscillation program. 
The agreement between these predictions and the data is comparable to the one of other theoretical models commonly used to calculate electron and neutrino cross sections, such as SuSAv2 and GiBUU. Results obtained with GENIE, a Monte Carlo event generator, are also  discussed for comparison. 
The good performances obtained with our neural network suggest that neural networks could be exploited for theoretical and experimental investigations of electron- and neutrino-nucleus scattering.

\end{abstract}

\maketitle

\section{Introduction}
For over a hundred years, the electron-scattering experiments have represented one of the most powerful and fruitful approaches for investigating the structure of physical systems. 
The Franck–Hertz experiment \cite{Franck_Hertz:1914}, which showed that the energy absorption by atoms is quantized by studying the flux of electrons through a vapor of mercury atoms can be considered as the progenitor of 
an important tradition of electron-scattering experiments allowing us to move the frontiers of our knowledge of the structure of matter: from atoms, to nuclei \cite{Hofstadter:1956qs}, to quarks, precisely discovered thanks to experiments \cite{Bloom:1969kc,Breidenbach:1969kd} of this kind. 

The possibility of using other lepton beams also appeared in parallel with the development of increasingly energetic and intense electron beams. In particular, starting from Pontecorvo's \cite{Pontecorvo:1959sn} idea and Schwartz's \cite{Schwartz:1960hg} project, experiments with neutrino beams have been performed, the first having led to the discovery of muon neutrino \cite{Danby:1962nd}.

Nowadays accelerator-based neutrino experiments, such as T2K \cite{Abe:2011ks} and NOvA \cite{Ayres:2007tu}, are performed for a precise determination of neutrino oscillation parameters. The next generation experiments DUNE \cite{Acciarri:2016crz} and Hyper-Kamiokande \cite{Abe:2015zbg} will play a central role in the neutrino oscillation program.  

To ensure the success of these experiments, a reduction in systematic errors to the level of a few percent is needed. Today, one of the greatest sources
of systematic errors are the neutrino–nucleus cross sections. In these experiments nuclear targets (such as C, O, and Ar) are involved, and in the energy
region of hundreds of MeV to a few GeV, these cross sections are known to a precision not
exceeding 20 \% \cite{Katori:2016yel,Alvarez-Ruso:2017oui}. Knowledge of these cross sections is crucial to determine the neutrino energy that enters the expression of the neutrino oscillation probability.
Since neutrino beams are not monochromatic, in contrast with electron beams, the initial neutrino energy is reconstructed from the final states of the neutrino-nucleus reaction.

The current program of electron-scattering experiments continues in connection with nuclear, hadronic and particle physics and represents a powerful tool in connection with neutrino physics. Indeed, electrons and neutrinos both being leptons, their interaction with atomic nuclei is similar: it happens via vector current in the case of electrons and via vector and vector-axial currents in the case of neutrinos. 
Hence the vector part of the cross section and the final state interaction of the scattered particles are identical for electron and neutrino scattering. Although the axial part of the neutrino cross section complicates the comparisons, 
the electron-scattering experiments, which have the great advantage of employing  monoenergetic electron beams, are useful in the investigation of  neutrino-nucleus scattering from several points of view.  First of all, the large amount of cross section data already obtained by many different electron-scattering experiments can be used to validate microscopic theoretical models employed to predict neutrino-nucleus cross sections; a non exhaustive list of such studies includes the works of Refs. \cite{Leitner:2008ue,Ankowski:2014yfa,Pandey:2014tza,Megias:2016lke}. These world data have been recently considered in Refs.  \cite{Ankowski:2020qbe,Barrow:2020mfy,electronsforneutrinos:2020tbf} to benchmark the neutrino Monte Carlo event generators, such as GENIE \cite{Andreopoulos:2009rq}, commonly used in the neutrino community, as well as to check the neutrino energy reconstruction via calorimetric and quasielastic kinematics-based methods \cite{CLAS:2021neh}. The available 
%database \cite{escat_webpage} 
data also includes the recently published electron cross sections on argon and titanium  \cite{Dai:2018xhi,Dai:2018gch,Murphy:2019wed,Gu:2020shc}. These cross sections were measured in the experiment proposed in Ref. \cite{Benhar:2014nca} to  determine the spectral function of $^{40}$Ar, the nucleus employed in DUNE, and in the detectors of the Fermilab Short-baseline neutrino program \cite{Machado:2019oxb}. 

The aim of this work is to investigate the use of neural networks to reproduce and predict the electron-nucleus cross sections in the kinematical domain of the present and future accelerator-based neutrino oscillation experiments. 

In the last decade, applications using neural networks have become ubiquitous and are found in many tasks beyond fundamental research. 
In the high energy physics community, deep learning provides a faster alternative to both standard data analysis techniques and Monte Carlo approaches to simulate the detector outputs \cite{Baldi:2014kfa, Albertsson:2018maf,Radovic:2018dip,Ju:2020xty,doi:10.1142/12200}. Deep learning for event reconstruction in accelerator-based neutrino experiments is used for example in Refs.  \cite{Acciarri:2016ryt,Perdue:2018ihs,Baldi:2018qhe,Abi:2020xvt,Alonso-Monsalve:2020nde,Abratenko:2020ocq,Ghosh:2021ytz}. 
Recent studies \cite{Alanazi:2020klf,Alanazi:2020jod}
applied generative adversarial networks 
to construct an AI-based Monte Carlo event generator
for deep-inelastic electron-proton scattering, free of theoretical
assumptions about the underlying particle dynamics. In the context of electron and neutrino scattering, theoretical studies have used neural networks to obtain information on the two main ingredients of the lepton-nucleus cross sections: nuclear responses, and nucleon form factors. 
More precisely, in Ref.\cite{Raghavan:2020bze} a physics-informed artificial neural network is employed to 
reconstruct the electromagnetic response functions. 
%from the Euclidean responses. 
In Ref. \cite{Alvarez-Ruso:2018rdx} a Bayesian approach for feed-forward neural networks has been applied to extract from the neutrino-deuteron scattering data the nucleon axial form factor, a  quantity which has been widely debated in the last ten years, following the MiniBooNE measurement of the quasielastic-like neutrino cross section on carbon \cite{AguilarArevalo:2010zc}. 
The possibility of using machine learning algorithms in reconstructing neutrino energy has been explored in Ref. \cite{Nagu:2021zho}. 
For a recent  review on the current trends and perspectives of artificial intelligence in nuclear physics we refer to \cite{Bedaque:2021bja}.

\section{Data}\label{section.Data}
A large amount of data is needed to predict electron-nucleus cross sections via a neural network approach. These data, accumulated by many different electron scattering experiments which began in the mid-1970s have been assembled by the authors of Refs.~\cite{Benhar:2006er,Benhar:2006wy} and made available in the ``Quasielastic Electron Nucleus scattering archive'' \cite{escat_webpage}. 

This archive contains about 600 different combinations of targets, energies and angles consisting of some 20000 data points. These data cover the energy region of interest for the present and future long-baseline neutrino oscillation experiments, from giant resonance excitations up to the deep-inelastic scattering, with a predominant contribution of quasielastic and $\Delta$ resonance excitations. 

\begin{table}
\centering
\begin{tabular}{| m{1cm} | m{1cm} | m{1cm}| m{1cm} | m{1cm} | m{1cm} | m{1cm} | m{1cm} | m{1cm} | m{1cm} | m{1cm} | m{1cm} | m{1cm} |} 
\hline
$^4$He & $^6$Li & $^9$Be & $^{12}$C & $^{16}$O & $^{24}$Mg & $^{27}$Al& $^{40}$Ar & $^{40}$Ca &  $^{48}$Ca & $^{48}$Ti & $^{56}$Fe & $^{59}$Ni \\ 
\hline
21.5   & 1.3  & 3.1   & 24.0     & 1.0     & 0.3     &  6.3 & 1.4 &  10.5  & 9.6  & 1.4 & 19.3 &  0.3  \\ 
\hline
\end{tabular}
\caption{Percent contribution of the different nuclear targets electron scattering data  \cite{escat_webpage} considered in our study.}
\label{table_nuclei}
\end{table}

The nuclei included in the archive \cite{escat_webpage} vary from hydrogen to uranium. 
Driven by the idea of having a relatively homogeneous data-set, containing the nuclei employed in neutrino detectors, we decided to discard the very light nuclei and the heavy nuclei characterized by a large neutron-proton asymmetry. The final subset of nuclei considered varies between $^4$He and $^{59}$Ni. 
They are specified in Table \ref{table_nuclei} together with their relative contribution to our dataset.

\begin{figure}
    \centering
\includegraphics[height=0.6\textheight]{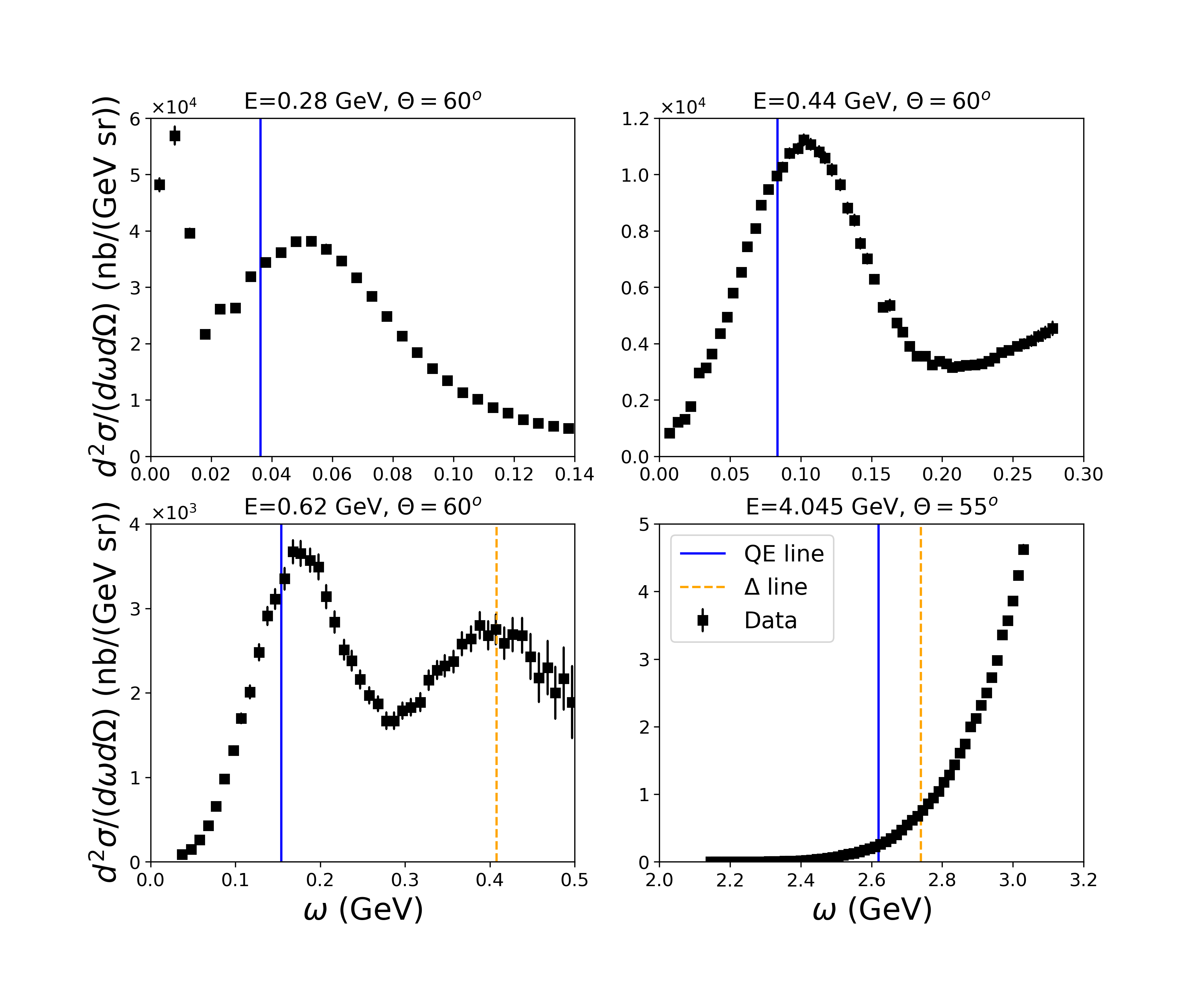}
    \caption{Inclusive $(e,e')$ $^{12}$C double-differential cross section data as a function of the transferred energy in four different kinematical conditions written above each panel. The vertical lines correspond to the kinematics of quasielastic
and $\Delta$ excitations given by Eq.(\ref{qe_line}) and Eq.(\ref{delta_line}), respectively. }
    \label{fig_4_plots_c12}
\end{figure}

In Fig. \ref{fig_4_plots_c12} we plot some examples of inclusive double-differential cross sections $\displaystyle \frac{d^2 \sigma}{d\omega d\Omega}$ for different values of incoming electron energy $E$ and electron scattering angles $\theta$\footnote{We recall that the relation between the differential solid angle $d\Omega$ in the direction specified by the scattered electron momentum and the electron scattering angle $\theta$ is $d\Omega=2\pi d\cos\theta$.} as a function of the energy transferred to the nucleus $\omega$ (also called energy loss, being $\omega=E-E'$ the difference between the incoming and outgoing electron energy $E'$). For this illustration we have selected $^{12}$C, a nucleus for which much data exists. Two characteristics that would complicate the neural network prediction task can already be seen from these few examples which refer to different kinematical conditions. Firstly, the cross section may span many orders of magnitude: 5 orders in the four panels of Fig. \ref{fig_4_plots_c12}, which refers to the same nucleus, and 12 orders of magnitude (from $4\times10^{-6}$ to $10^6$) when considering the whole dataset. 
Secondly, the shapes of the cross section can vary greatly, reflecting the different reaction mechanisms that can be induced by different kinematics. Starting from the top left panel of Fig. \ref{fig_4_plots_c12}, the first recognizable excitations are the nuclear giant resonances which correspond to the sharp peaks in the cross sections for $\omega<20$ MeV. Moreover, the quasielastic bump appears in the three first panels. It corresponds to one nucleon knockout and 
is peaked around 
\begin{equation}
\label{qe_disp_rel}
\omega_{QE}=\frac{Q^2}{2M_N}=\frac{{\bf{q}}^2-\omega^2}{2M_N}=\sqrt{{\bf{q}}^2+M_N^2}-M_N, 
\end{equation}
where $M_N$ is the nucleon mass and ${\bf{q}}$ the momentum transfer to the nucleus, given by the difference between the incoming and scattered electron momentum,  ${\bf{q}}={\bf{k}}-{\bf{k'}}$. In terms of electron kinematics variable, when electron mass is neglected, Eq.(\ref{qe_disp_rel}) can be written as: 
\begin{equation}
\label{qe_line}
\omega_{QE}=\frac{{E}^2 (1-\cos\theta)}{M_N + E (1-\cos\theta)}.
\end{equation}
A vertical line corresponding to this value is plotted in each panel of Fig. \ref{fig_4_plots_c12}. The shift of the position of the real quasielastic peak with respect to the value of Eq.\eqref{qe_disp_rel} and Eq.\eqref{qe_line} is due to the absence of nucleon binding and nuclear collective effects in these formulae. The broadening of the quasielastic bump is due to nucleon Fermi motion.  
The second bump at a higher $\omega$ corresponds to the $\Delta$ resonance excitation. In the case of scattering with a free nucleon at rest, it would peak at:
\begin{equation}
\label{delta_disp_rel}
\omega_{\Delta}=\sqrt{{\bf{q}}^2+M_{\Delta}^2}-M_N=\frac{Q^2}{2M_N}+  \Delta  M, 
\end{equation}
with $ \Delta  M =(M^2_{\Delta}-M^2_N)/2M_N= 338$ MeV. Equivalently, Eq.\eqref{delta_disp_rel} can be written as: 
\begin{equation}
\label{delta_line}
\omega_{\Delta}=\frac{M_N{ \Delta  M +E}^2 (1-\cos\theta)}{M_N + E (1-\cos\theta)}, 
\end{equation}
also shown in Fig. \ref{fig_4_plots_c12}. 
Beyond the $\Delta$ excitations, higher energy nucleon resonances contribute to the cross section up to the onset of the  deep inelastic scattering. On the other hand, in the $\omega$ region between the quasielastic and $\Delta$  peaks, the so called ``dip'' region, a large part of the cross section is due to multinucleon excitations, arising from nucleon-nucleon correlations and meson exchange processes. These complicated many-body mechanisms have attracted a lot of attention in the neutrino community in recent years, starting from the suggestion \cite{Martini:2009uj} of their inclusion   as natural explanation of the MiniBooNE cross sections \cite{AguilarArevalo:2010zc}. These mechanisms play a crucial role in the neutrino energy reconstruction problem \cite{Martini:2012fa,Martini:2012uc,Nieves:2012yz,Lalakulich:2012hs,Ankowski:2016jdd}.  

In %our experiments, 
this work, 
we remove the  small number of data points corresponding to the low-energy giant resonances. Their presence worsens the neural network predictions for essentially two reasons: they are characterized by the largest (and spiked) cross sections, and they represent a small proportion of the whole dataset, which focuses on quasielastic excitations and beyond. %These excitations are beyond the scope and this work though we intend to include them in future work. 
We postpone the inclusion of these excitations to future work.

Several microscopic quantum-mechanical and phenomenological approaches exist to describe the different reaction mechanisms that contribute to the inclusive electron-nucleus cross section. While some approaches allow a description of all the nuclear excitations mentioned earlier, others focus on specific excitations only. For a review of these models we refer the reader to Ref. \cite{Benhar:2006wy} or to Ref. \cite{Amaro:2019zos}, where different approaches are discussed in connection with both electron and neutrino scattering. In the following we analyze to what extent neural networks can reproduce the electron-nucleus cross sections. 

\section{Neural Networks}

 In regression tasks, machine learning models are designed to learn mappings between an input domain and a predefined set of outputs for the target variables. 
 The dataset considered in this work is composed of different features: the atomic number $Z$,  the nucleus mass number $A$, the electron beam energy $E$, the electron scattering angle $\theta$, the energy loss $\omega$ and the inclusive double-differential cross section $\displaystyle \frac{d^2 \sigma}{d\omega d\Omega}$. This last quantity is the value to be predicted by the supervised learning algorithm we present. 
Beyond the five data features ($Z$, $A$, $E$, $\theta$, and $\omega$), we add four additional complementary variables, obtained as combinations of the five original ones. 
These added variables are: $\cos \theta$, $\omega_{QE}$ defined in Eq.(\ref{qe_line}), $\omega_{\Delta}$ defined in Eq.\eqref{delta_line}, and $Q^2={\bf{q}}^2-\omega^2$, here defined as:

\begin{equation}
\label{E_omega}
Q^2
=2E(E-\omega)(1-\cos \theta),
%=E^2+(E-\omega)^2-2E(E-\omega)\cos(\theta)- \omega^2
\end{equation}
a relation that is only valid when the electron mass is set to zero.
 
The addition of handcrafted features, such as the four-momentum transfer in the reaction and the approximate position of the quasielastic and $\Delta$ peaks 
(despite a possible offset between real and approximated values due to the lack of removal energy parameters in Eqs. (\ref{qe_line}) and \eqref{delta_line}), is found to be useful to drive the network.

A larger neural network is expected to perform equally well without adding handcrafted features but our choice to include these additional variables makes the optimization process easier and allows us to use a smaller network, that is faster to train and less prone to overfitting. However, exactly quantifying how these four variables actually improve the prediction is not possible since there is no guarantee that the fine tuning is equally well performed when we add the four additional variables and when we do not add them.

The model is built upon deep neural networks (DNNs) \cite{Goodfellow:2015}. DNNs are designed to learn hierarchical and increasingly abstract representations of the data. A DNN is a composition of $L$ parametric functions named layers. The output of each layer, $f_l, l \in \{0,\ldots ,L-1\}$, is understood as a representation of the input samples. More specifically, the layer is composed of neurons, which are the building blocks of a layer. 
Hence, the layer $f_l$ takes the output of the previous layer $f_{l-1}$ and applies a non-linear transformation to compute its output. These transformations use the model parameters, $W_l$ for each layer, commonly called \textit{weights} and conveniently represented by a rectangular matrix. They relate the neurons of a layer to the previous layer and contain the information extracted by the model from the training data. Thus, given an input $x$, a neural network $f$ performs the following computation to infer its output:
\begin{equation}\label{eq:neural_net}
    f(W,x) = f_{L-1}(W_{L-1}, f_{L-2} (W_{L-2}, \ldots , f_{0} (W_{0},  x)) \ldots ).
\end{equation}

In Eq. (\ref{eq:neural_net}), the state of layer $l$ can be denoted $x^l$ and is called a representation. In a standard feedforward network, $x^{l-1}$ is employed to compute $x^l$ as follows:
\begin{equation}\label{eq:layer_neural_net}
x^l=f_{l-1}(W_{l-1},x^{l-1})=g(W_{l-1}*x^{l-1}+ b_{l-1}),
\end{equation}
where $W_{l-1}*x$ is a matrix product, $x$ can be seen as a column vector of $n_{l-1}$ components, then $W_{l-1}$ is a matrix of $n_{l-1}$ columns and $n_{l}$ lines, $b_{l-1}$ is a column vector of $n_{l}$ lines. $g$ is an element-wise nonlinear function. We used the standard Rectified Linear Unit, ReLU function defined as $\text{ReLU}(z)=\text{max} \{0,z\}$. Other choices might work equally well or even better, we make no claim of optimality in this paper. The result of Eq. (\ref{eq:layer_neural_net}) is the input of the next layer of the neural network, layer $l$. 

During the initial exploratory phase of this work we used fully connected neural networks as described by Eq. (\ref{eq:layer_neural_net}). We found that the richness of the data was such that deep neural networks with less than 35 layers were unable to fit the training set. We therefore used deeper networks that then became difficult to train for the reasons explained in Ref.  \cite{He:2015Resnet}. The solution proposed by the authors of Ref. \cite{He:2015Resnet} is to use what they named \textit{a residual network}. The architecture of our neural network is a fully connected \textit{residual network} \cite{He:2015Resnet} which means that we perform an addition that shortcuts the architecture to help the optimization procedure converge. The skipped connection is performed as follows:
\begin{equation}\label{eq:skip_layer}
x^l=f_{l-1}(W_{l-1},x^{l-1},x^{l-5} )=g(W_{l-1}*g(x^{l-1}+x^{l-5})+ b_{l-1}).
\end{equation}

\begin{figure}[ht]
    \centering
\includegraphics[height=0.5\textheight]{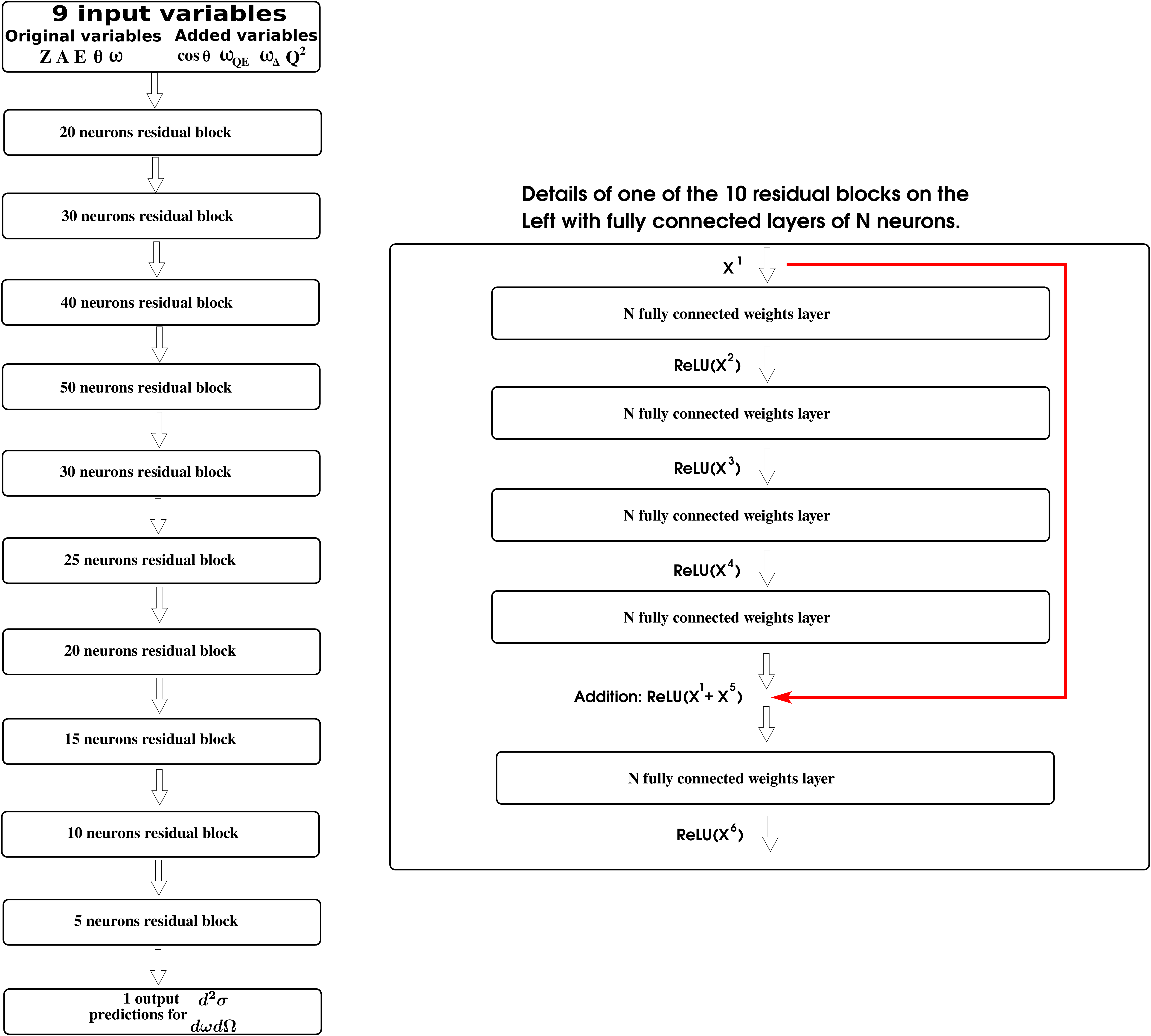}
    \caption{Illustration of the chosen neural network, composed of ten fully connected residual blocks. Details of one of these residual blocks are shown on the right of the figure. The activation function used is a Rectified linear unit (ReLU).}
    \label{figNeuralNet}
\end{figure}

The skipped connections or residual blocks are used in deep learning to help training deep neural networks \cite{Goodfellow:2015,He:2015Resnet}. 
Our neural network, implemented with 
%\sout{Keras \cite{chollet2015keras}}
TensorFlow \cite{tensorflow2015-whitepaper}, is composed of 10 residual blocks of 5 fully connected layers with increasing then decreasing sizes ranging from input size (9 input data) up to size 50 then down to the size 5 before connecting to the output of size one. Our neural network architecture is represented in Fig. \ref{figNeuralNet}, the resulting model has 50 connected layers. To help information flow in the backpropagation stage we have chosen a layer width that varies progressively. Abrupt changes in widths made the optimization process harder and local minima were difficult to escape from. The constraint we have is that we start with 9 inputs and have an output of dimension 1.

An essential element of neural networks training is the choice of the cost function. For the neural network to be able to accurately predict the cross section structure over its %\sout{many orders of magnitude} 
wide range and different shapes, we chose to use the relative absolute error given by: 
%$\displaystyle
\begin{equation}
    C(\hat{y})=\frac{1}{n}\sum_{i=1}^n \frac{|y_i-\hat{y}_i|}{y_i}.%, %$,
    \label{eq_cost}
\end{equation}
In Eq. (\ref{eq_cost}), $\hat{y_i}$ is a model prediction and $y$ is the cross section divided by $A$, $y=\displaystyle \frac{\frac{d^2 \sigma}{d\omega d\Omega}} {A}$ obtained from the dataset. In our case, 
\begin{equation}\label{eq:neural_net_pred}
    \hat{y_i}=f(W,x_i), 
\end{equation}
where the input data is the nine dimensional vector of normalized components : $x_i=(Z, A, E, \theta, \omega, \cos\theta, \omega_{QE}, \omega_{\Delta} ,Q^2 )$ and $(x_i, y_i)$ is an input-output pair. Another choice for the loss function could be the $\chi^2$ that would allow to take into account the experimental errors. For this first exploratory study, driven by a major simplicity and rapidity of calculations and by the fact that in the present data set the experimental errors are in general small,  we prefer to chose the relative absolute error as cost function. We leave the use of $\chi^2$ to future works. 

During the training phase, the model is presented with a large number of input-output pairs. The weights, $W$, are initialized randomly with a normal distribution \cite{Goodfellow:2015} and an initial prediction of the output is computed using Eq.\eqref{eq:neural_net}. During the forward pass, the model’s prediction error is estimated by computing the value of a cost function that quantifies the discrepancy between the current prediction $\hat y$ given by Eq. \eqref{eq_cost} and the true target $y$. This cost function is computed over a batch of data and differentiated with respect to the model weights, $W$, in the backward pass. Model weights are then updated with the computed differential to improve the predictions of the neural network. The general idea is to update $W$ in the following way: 
\begin{equation}
    W=W-\epsilon \frac{\partial C}{\partial W}.
    \label{eq_update}
\end{equation}
The variable $\epsilon$ is called the \textit{learning rate} and the value we chose was standard, $\epsilon=0.001$. In practice, we use Adam optimizer \cite{kingma2014method} which provides an empirical improvement over Eq. \eqref{eq_update} with a slowly decreasing learning rate, reminiscent of annealing, to help the optimization process. The values of the neural network weights that minimize the cost function are obtained during the training phase by iteratively taking forward and backward passes on the training data set.
Once a predetermined number of \textit{epochs} (a forward and a backward pass) is computed, the training phase is finished and the model is deployed on new samples, unseen during the training, to make predictions. Training takes about 30 minutes for 5000 epochs on a 12GB NVIDIA Tesla K80 GPU. 
We include standard $L_2$ regularization in order to reduce the variance of the model and hopefully improve \textit{generalization}: the predictions on data not used during the training phase. Weights are stored during training and for inference we roll back to the epoch that yielded the lowest validation error during the whole training. In the inference phase, the output of the model is computed for the test data set. One hopes that the model will generalize well on the test data and infer target values close to the actual values thanks to the patterns it has learned from its training data.

\section{Results}
The training set is composed of the available samples excluding one or more chosen isotopes $(Z',A')$, or examples corresponding to some particular kinematical conditions. These excluded data will be examined in the testing stage. 
In other words, we used the trained neural network to make predictions for the isotopes $(Z',A')$ or for some specific kinematical conditions. This allows us to assess the ability of the trained neural network to infer the desired target for cases unseen during the training stage.
 
\subsection{Training results}

\begin{figure}
    \centering
\includegraphics[height=0.3\textheight]{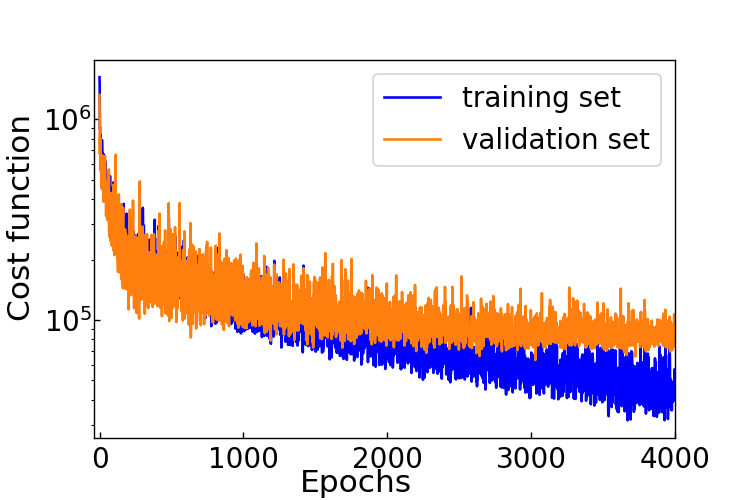}
    \caption{Evolution of the cost function during the 4000 epochs training process for a successful training run. This illustrates a slight overfitting, controlled by $L_2$ regularization.}
    \label{figLossEvol}
\end{figure}

To further illustrate our procedure let us consider some specific cases. The whole dataset, representing our starting point, is the subset of the electron scattering data \cite{escat_webpage} corresponding to nuclei from $^4$He to $^{59}$Ni. Their relative contribution is illustrated in Table \ref{table_nuclei}. As a first experiment, we remove from the entire dataset only the isotope $(Z=20, A=40)$, \textit{i.e.} the $^{40}$Ca, our test nucleus, for which we intend to predict the cross section. The rest of the dataset is randomly split: $90\%$ for the training set and the remaining $10\%$ for the validation set that allows us to monitor overfitting.

Stochastic gradient descent on minibatches is used for parameters optimization \cite{Goodfellow-et-al-2016}.
The minibatches of $512$ instances of the training set are constituted randomly. This promotes regularization and speeds up the training. As a standard practice, at the end of the training process over 4000 epochs, we plot the evolution of the cost function, displayed in Fig. \ref{figLossEvol}, to check that the training has converged as expected and to monitor overfitting. In order to reduce the variance of the model and to control overfitting we used $L_2$ regularization on the first and largest layers. It is standard practice to regularize large layers, that are more prone to overfitting, more than small layers. To limit overfitting we use for our prediction the parameters that gave the best validation value. 

\begin{figure}
    \centering
\includegraphics[height=0.85\textheight,
width=1.0\textwidth]{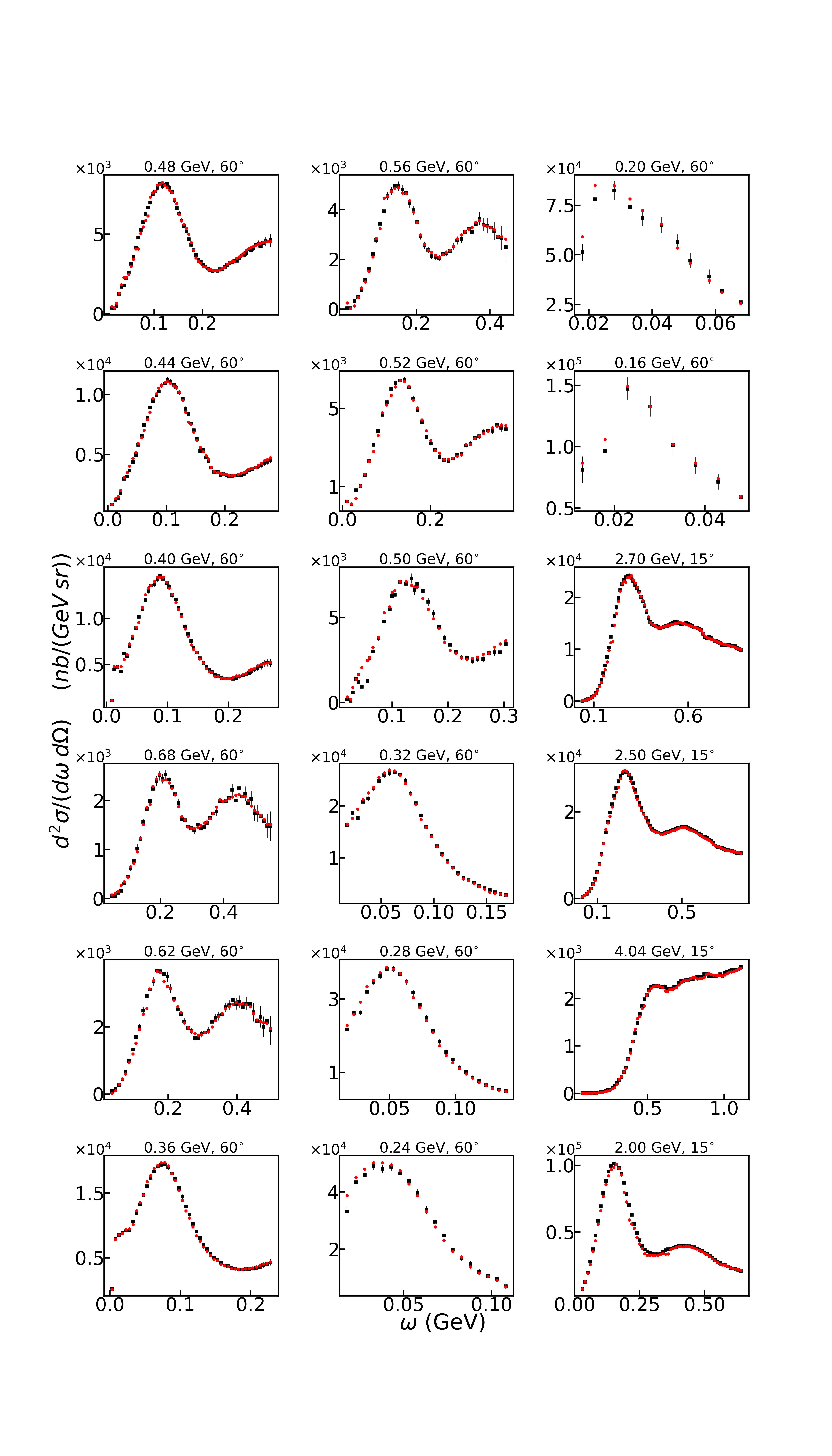}
    \caption{Comparisons between the experimental inclusive $(e,e')$ double differential cross section on $^{12}$C, in black squares with experimental error bars, and the neural network's predictions, red circles, for the different values of incoming electron energy $E$ and scattering angle $\theta$ (indicated above each panel), as a function of the transferred energy. To obtain these predictions, the $^{12}$C data occurrences were used in the training set as well as all the other nuclei of the dataset except $^{40}$Ca.}
    \label{figC}
\end{figure}

\begin{figure}
    \centering
\includegraphics[height=0.85\textheight,
width=1.\textwidth]{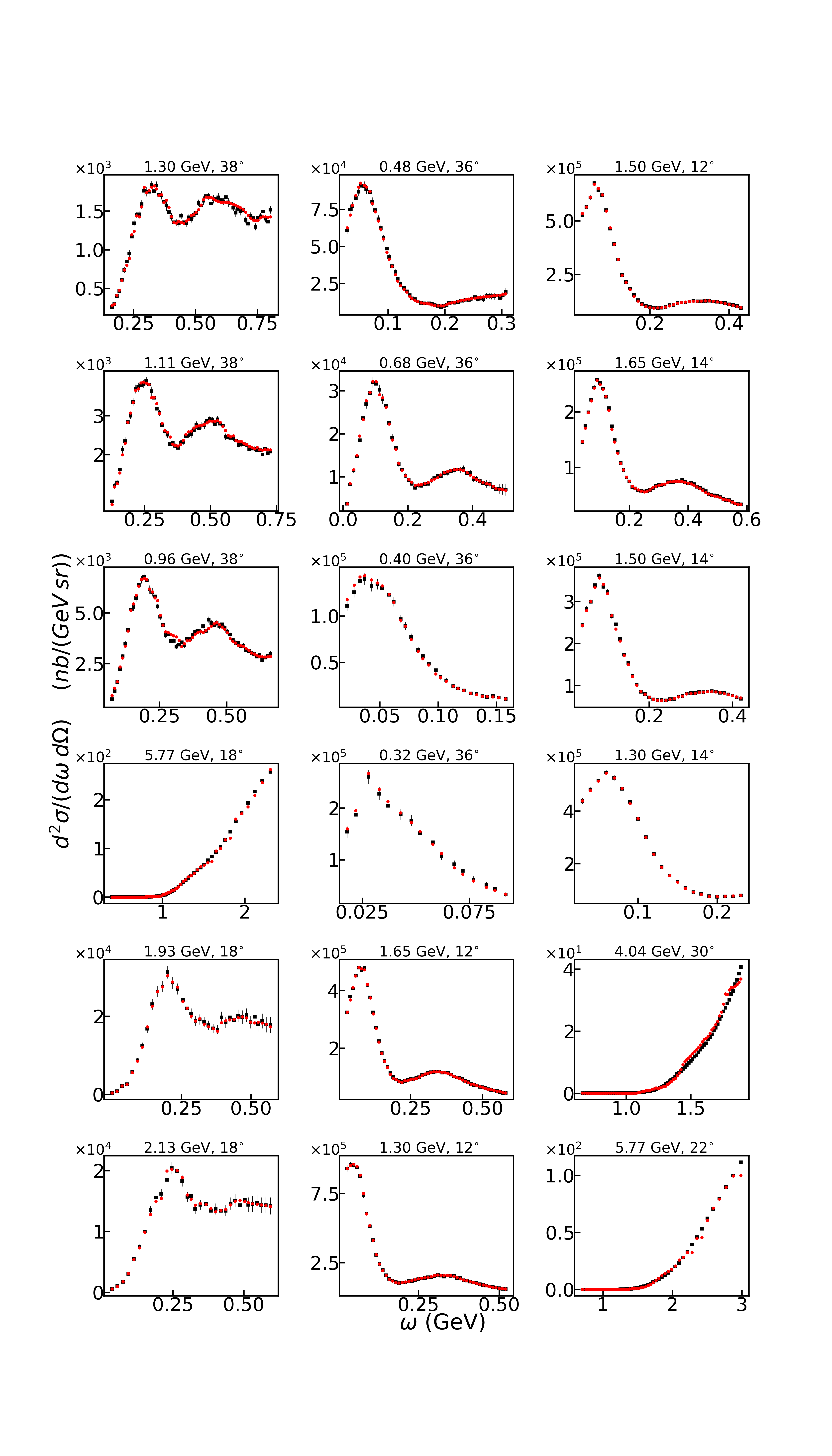}
    \caption{The same as Fig. \ref{figC} but for other kinematical conditions.}
    \label{figC_2}
\end{figure}

In order to illustrate the overall performance of the neural network on the training set, we consider its predictions for $^{12}$C,
a nucleus currently employed in the T2K and NOvA neutrino detectors. For this nucleus, much electron-scattering cross section data are available, for different kinematical conditions, inducing different nuclear excitations, from nuclear giant resonances up to deep inelastic scattering, as discussed in Section \ref{section.Data}. 

The neural network double-differential cross section results for $^{12}$C are illustrated in Figs. \ref{figC} and \ref{figC_2} for a selected set of the values of incoming electron energy $E$ and electron scattering angle $\theta$ present in the database. 
The overall agreement with data is remarkable for all the incoming energies and electron scattering angle\footnote{There are some additional kinematical conditions that are not shown in Figs. \ref{figC} and \ref{figC_2} to avoid overwhelming figures but the agreement between the data and the neural network fit on the training set remains remarkable also in the cases not shown.}.

These good results suggest that neural networks can be considered as a tool for representing data that has this kind of behavior (changing shape depending on the type of induced excitation). In this context, we recall that for $^{12}$C  the values of $\frac{d^2 \sigma}{d\omega d\Omega}$ vary between a minimum of $7.7 \times 10^{-5} \frac{\textrm{nb}}{\textrm{GeV sr}}$ and a maximum of $9.6 \times 10^{5} \frac{\textrm{nb}}{\textrm{GeV sr}}$. Despite this variation of $10$ orders of magnitude, the choice of the mean relative absolute error of Eq.\eqref{eq_cost} as cost function allows for efficient training of the neural network. 

\subsection{Testing results} 
\begin{figure}
    \centering
\includegraphics[height=0.85\textheight,width=.85\textwidth]{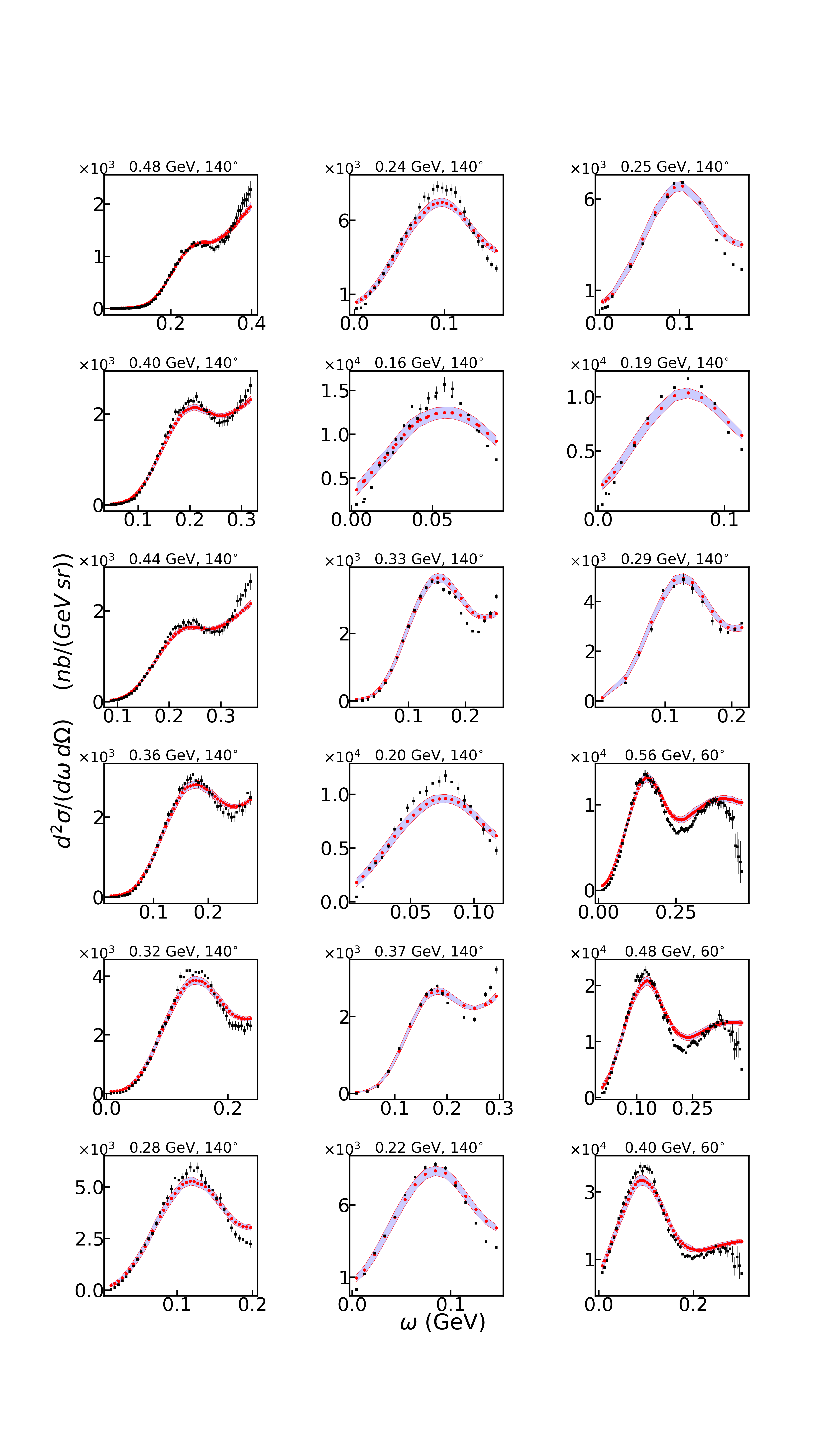}
    \caption{Comparisons between the experimental inclusive $(e,e')$ double-differential cross section on $^{40}$Ca, in black squares with experimental error bars, and the neural network's predictions, red circles, for the different values of incoming electron energy $E$ and scattering angle $\theta$ written above each panel, as a function of the transferred energy. These are the predictions for the test data for which the neural network has not been trained on. $95\%$ confidence interval is shown by the shaded blue area, bounded by red lines.}
    \label{figCa40}
\end{figure}

\begin{figure}
    \centering
\includegraphics[height=0.85\textheight,width=.85\textwidth]{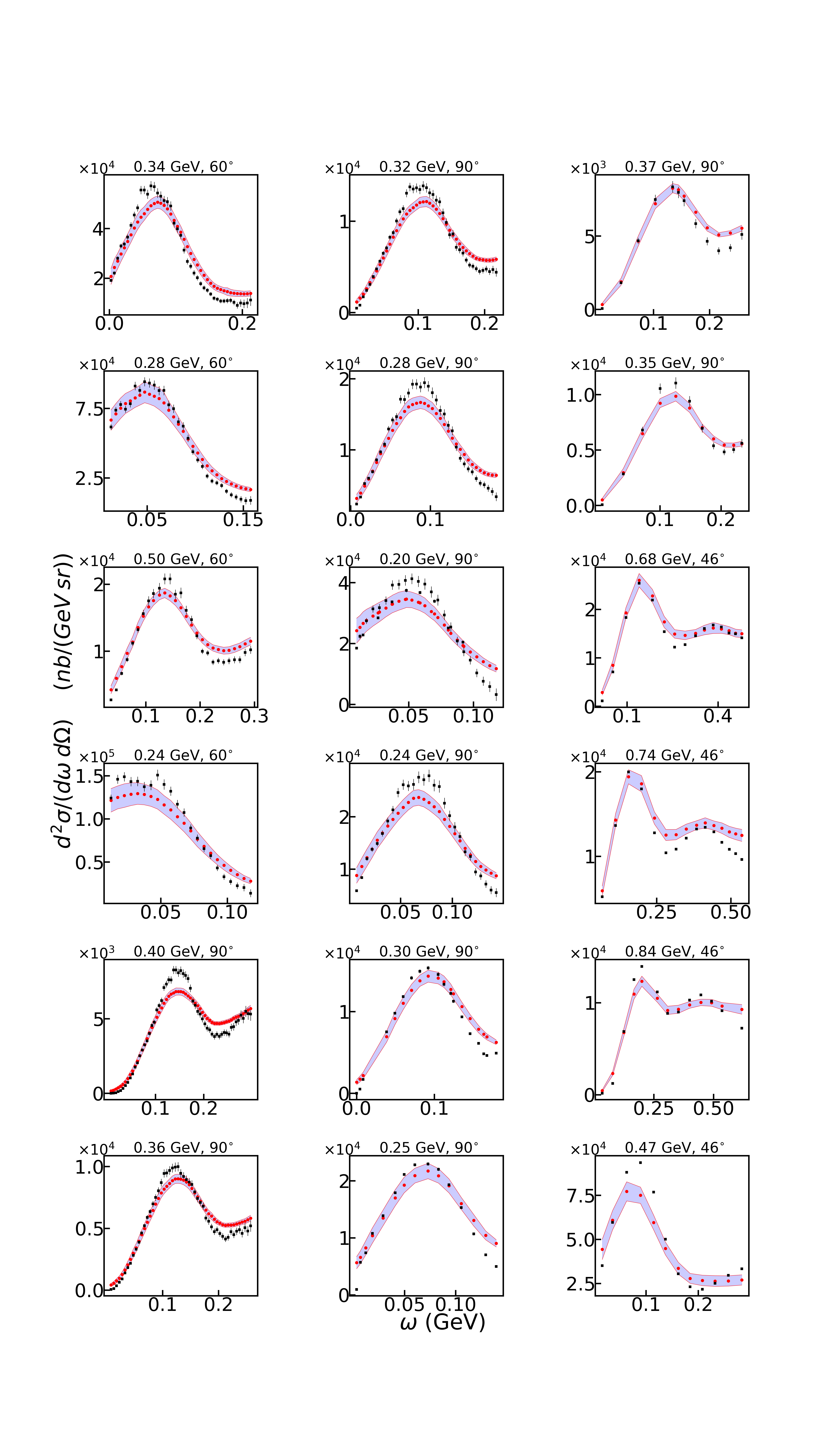}
    \caption{The same as Fig. \ref{figCa40} but for other kinematical conditions.}
    \label{figCa40_2}
\end{figure}

Following the results obtained for the training set, in this section we investigate the predictive power of the neural networks for the test set. 
First, we consider $^{40}$Ca as test nucleus. This choice is motivated by the existence of many available cross section data for this nucleus. The neural network predictions for this nucleus, which, we reiterate, was not included in the training set, are shown in Figs. \ref{figCa40} and \ref{figCa40_2}. These figures show the result of an average over $N=100$ different neural networks trained on the same dataset. After initializing the same neural network architecture with $N$ different realizations of a random variable, we train the neural networks. For each training run we keep the weights that performed best on the validation set. These different neural networks give different predictions on the test set. For each scattering angle $\theta$, each energy $E$, and each transferred energy $\omega$, we obtain $N$ independent values for $\hat{y}_i$, the prediction for $\frac{d^2 \sigma}{d\omega d\Omega}$. We can then estimate the average based on these $N$ different predictions. Assuming a Gaussian distribution of these predictions, average and $95\%$ confidence intervals for the true average $\mu$ are estimated as follows: 
\begin{align}
\overline{y}=&\frac{1}{N} \sum_{i=1}^{N} 
\hat{y}_i  \\
\sigma^2=&\frac{1}{N-1} \sum_{i=1}^{N} \left ( 
\hat{y}_i -\overline{y}
\right)^2 \\
&|\overline{y} - \mu | < 1.96\times \frac{\sigma}{\sqrt{N}}.
\end{align}

The estimated average of the predictions is plotted in Figs. \ref{figCa40} and \ref{figCa40_2} together with the estimated $95\%$ confidence interval, bounded by two red lines and shaded in blue. One can observe that the overall agreement of our neural network predictions with the data for $^{40}$Ca is satisfactory from quasielastic up to the $\Delta$ resonance. The prediction results could be further improved by adding more examples into the training set, with different choices of the neural-network hyperparameters or including more handcrafted features. One could also use alternative approaches, like Bayesian neural networks to better take into account the uncertainty. However, for the exploratory purpose of the present work, we do not claim optimality in any sense, and our focus is on presenting the interest of the method.

As an additional illustration of the predictive performances of the neural network approach we consider some tests on nuclei of particular interest for the neutrino-oscillation program. Namely oxygen, employed in the present and future water Cherenkov detectors Super-Kamiokande and Hyper-Kamiokande, and argon, the nucleus employed in three detectors of the present short-baseline Fermilab neutrino programs as well as by the future DUNE experiment. 

\begin{figure}
\begin{center}
%    \centering
\includegraphics[width=1.0\textwidth]{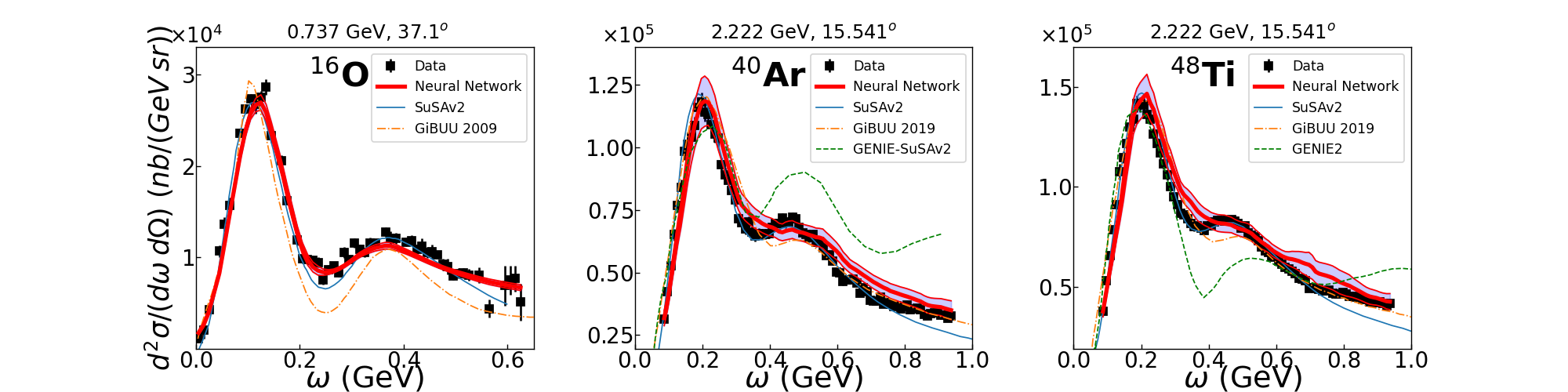}
%{Pannel_O_Ar_Ti_4_7_22.png}
    \caption{
 Comparison between the inclusive $(e,e')$ double differential cross section  data and the neural network's predictions for $^{16}$O (left panel), $^{40}$Ar (middle panel) and $^{48}$Ti (right panel). The different values of incoming electron energy and scattering angle are written above each panel. For the left panel, the neural network was trained without the data represented on the figure. For the middle and right panels, the neural network was trained without both $^{40}$Ar and $^{48}$Ti data.  Theoretical predictions of SuSAv2 and GiBUU approaches are also shown. They are taken respectively from Refs.\cite{Amaro:2019zos} and \cite{Leitner:2008ue} for $^{16}$O and from Refs. \cite{Barbaro:2019vsr} and \cite{Mosel:2018qmv} for $^{40}$Ar and $^{48}$Ti. GENIE Monte Carlo event generator results, taken from Ref. \cite{electronsforneutrinos:2020tbf} in the case of $^{40}$Ar and from Ref. \cite{Ankowski:2020qbe} in the case $^{48}$Ti, are also displayed. The red lines are the boundary of the $95\%$ confidence interval shaded in blue. }
    \label{Pannel_O_Ar_Ti}
    \end{center}
\end{figure}

In order to investigate the predictive power of our neural network for some specific kinematical conditions, in the case of oxygen (for which only few data are available, as shown in Table  \ref{table_nuclei}) we remove from the training data set all the events corresponding to electron scattering on $^{16}$O with a scattering angle of $\theta= 37.1^\circ$ and an incoming electron energy of $E=0.737$ GeV. The predictions are shown in the left-hand panel of Fig. \ref{Pannel_O_Ar_Ti}. On the same figure, in its middle and right-hand panel, we show the neural network predictions for argon and titanium. 
For these two nuclei, the inclusive electron-scattering data has been made available only recently \cite{Dai:2018xhi,Dai:2018gch,Murphy:2019wed}. Due to the increasing interest in argon detectors, the results of titanium ($Z=22$) represent an important source of information for the charged current interaction of neutrinos with a neutron of $^{40}$Ar ($Z=18, N=22$). For $^{40}$Ar and $^{48}$Ti, experimental data only exist for one particular experimental set-up (the same for both nuclei), corresponding to $E=2.222$ GeV and $\theta=15.541^\circ$. We removed from the usual dataset data points corresponding to these two nuclei in order to avoid the risk of overfitting due to the presence in the training dataset of the other similar nucleus, with the same kinematical conditions. For all three panels of Fig. \ref{Pannel_O_Ar_Ti}, the represented $95\%$ confidence interval is computed by considering $N=40$ independent neural network predictions, following the  methodology previously explained for $^{40}$Ca (where it was taken $N=100$). In the case of $^{16}$O the band is very narrow because the confidence intervals are small but in the case of $^{40}$Ar and $^{48}$Ti it is more visible in the figure. We explain this because of the complete absence of $^{40}$Ar and $^{48}$Ti in the training set.

For all three cases shown in Fig. \ref{Pannel_O_Ar_Ti}, the neural network predictions are excellent for the whole data set, from the quasielastic bump to the $\Delta$ resonance, passing by the ``dip'' region. Similar agreement has also been obtained by other theoretical approaches, such as SuSAv2 \cite{Amaro:2019zos} and GiBUU \cite{Buss:2011mx}, largely employed in neutrino cross section studies. 
We briefly recall the reader that SuSAv2, an updated version of the Super Scaling Analysis (SuSA) \cite{Amaro:2004bs} approach, is based on the observation of the superscaling behavior (\textit{i.e.} on the simultaneous $q$ and Fermi momentum independence) of the scaling function (a sort of nuclear response function in terms of a scaling variable) extracted from electron scattering data. Similarly to the neural network, it can be considered as a phenomenological data-driven approach, obviously with a much lower number of parameters and with a sound physical basis which microscopically justify it \cite{Gonzalez-Jimenez:2014eqa}. 

The Giessen Boltzmann-Uehling-Uhlenbeck (GiBUU) framework is an implementation of transport theory which allows a description of many nuclear reactions. It allows consistent treatment of the initial reaction vertex and of the final state processes.  It takes into account various nuclear effects via the local density approximation for the nuclear ground state, mean-field potentials, and in-medium spectral functions. In the implementation of Ref. \cite{Leitner:2008ue}, from which we take the electron cross section results of oxygen shown in Fig. \ref{Pannel_O_Ar_Ti},  multinucleon excitations were not included, which explains the GiBUU's underestimation of data in the ``dip'' region in this case. The underestimation was reduced in the GiBUU calculations for argon and titanium \cite{Mosel:2018qmv} by using an updated version of the model that takes into account the multinucleon contributions. We note that the ``dip'' region, which is particularly difficult to microscopically describe and reproduce with theoretical approaches, is well predicted by the neural network.  

For completeness, in Fig. \ref{Pannel_O_Ar_Ti} we also plot the results obtained with two different versions of GENIE (one of the most commonly used Monte Carlo event generators in the neutrino community) for $^{40}$Ar and $^{48}$Ti, taken from Refs. \cite{electronsforneutrinos:2020tbf} and \cite{Ankowski:2020qbe}, respectively. One can observe that starting from the ``dip'' region GENIE results disagree with the data. This disagreement is analyzed in  Refs. \cite{Ankowski:2020qbe} and \cite{electronsforneutrinos:2020tbf}, where several results are also shown and discussed for inclusive electron scattering on carbon.

\section{Summary and perspectives}

We have deployed a neural network model to predict the electron scattering inclusive double-differential cross sections on nuclei. The neural network predictions have been compared with a large amount of electron scattering data collected in the past as well as with recent data on Argon and Titanium that are of interest to the neutrino community. Our results show that neural networks can reproduce and predict the electron scattering cross section with an accuracy comparable to that provided by the microscopic approaches developed in the past and nowadays generalized to investigate the neutrino-nucleus scattering.

Neural network approaches can be impaired by a limited amount of data. Therefore, it was not a given that the data available was sufficient to train the model to make inference for unseen data. The many orders of magnitude spanned by the cross sections and their varying shape due to the different types of nuclear excitations (quasielastic, multinucleon, resonance excitations, and deep-inelastic) induced in different scattering kinematics make inference challenging. The neural network performance suggests that it could be used as an additional tool in the studies of electron and neutrino scattering on nuclei. Moreover, this can be used to predict the electron cross sections for nuclei and/or for  kinematical conditions where experimental data are absent and employ these predictions to validate Monte Carlo simulations. It could also be employed as a support to drive and speed up microscopic evaluation of cross sections and/or response functions, in some sense by generalizing what has been performed in Ref. \cite{Raghavan:2020bze} beyond the quasielastic excitation considered in that work. 

Another perspective involves the use of neural networks to directly predict the neutrino-nucleus cross sections. This task is not trivial since  the neutrino beams are not monochromatic. So in this case the measured quantity is the flux-integrated double differential cross section in terms of the final state measurable variables. For charged-current scattering process $(\nu_l, l)$ on nuclei, where $l$ is the charged lepton, this neutrino cross section reads:
\begin{equation}
\label{cross}
\frac{d^2 \sigma_\nu}{dE_{l}d\Omega}=
\frac{1}
{ \int \Phi(E_{\nu_l})~d E_{\nu_l}}
 \int d E_{\nu_l}
\left[\frac{d^2 \sigma_\nu}{d \omega  d\Omega}\right]_{\omega=E_{\nu_l}-E_{l}} \Phi(E_{\nu_l}), 
\end{equation}
where $E_l$ is the charged lepton energy  
and $d\Omega$, the differential solid angle in the direction specified by the charged lepton momentum. They represent the two final state measured variables. This expression formally reduces to that analyzed in this paper, when the flux of incoming particles, $\Phi$, reduces to a delta distribution. This is the case of monochromatic electron beams. For neutrino scattering, the problem is more complex since 
for a given set of the measured variables $E_l$ and $\Omega$,
one explores the full energy spectrum of neutrinos above the charged lepton energy, being $E_{\nu_l}=E_l+\omega$. As a consequence, all the reaction channels (giant resonances, quasielastic, multinucleon excitations, pion production arising from nucleon resonances decay, and deep-inelastic scattering)
are entangled and isolating a primary vertex process from the measurement of
neutrino flux-integrated differential cross section is much more difficult. 
When employing a deep learning approach for neutrino scattering, the technical challenge would be to design a network sufficiently rich to encode the complexity of the cross section for different primary vertex processes over the phase space relevant to the signal process. A similar approach is currently under study in the high energy collider community \cite{Albertsson:2018maf} in connection with the so called ``matrix element method''. This possibility should be investigated with the perspective of the development of an AI-based Monte Carlo event generator for neutrino-nucleus scattering.

\section*{Acknowledgements} 

We would like to thank Fabrice Couderc for his precious suggestions on neural network optimisation. We warmly thank Stefano Fortunati for interesting conversations on statistical topics.
We also acknowledge Christophe Coud\'e, Rapha\"el Lasseri, Gabriel Perdue, Pierre-Etienne Pion, David Regnier, Guillaume Scamps and   Kazuhiro Terao for fruitful discussions and exchanges on neural networks in connection with nuclear and neutrino physics. We thank Gwendolen Rodgers for proofreading the English. 

%\newpage
\bibliography{bib_micro,bib_Neuralnet}
%\bibliography{bib_Neuralnet}
\end{document}